\documentstyle[12pt]{article}

\newcommand{\be}{\begin{equation}}
\newcommand{\ee}{\end{equation}}
\newcommand{\bea}{\begin{eqnarray}}
\newcommand{\eea}{\end{eqnarray}}

\def\pa{\partial}
\def\tr{{\mathrm Tr\,}}
\def\G{{\cal G}}

\def\M{{\cal M}}
\def\H{{\cal H}}

\def\F{{\cal F}}

\def\R{{\mathbf R}}

\def\Ad{{\mathrm Ad\,}}
\def\ad{{\mathrm ad }}

\begin{document}

\thispagestyle{empty}
\setcounter{page}{0}

\renewcommand{\thefootnote}{\fnsymbol{footnote}}
\fnsymbol{footnote}

\vspace{.4in}

\begin{center} \Large\bf 
The Chiral WZNW Phase Space \\

and its Poisson-Lie Groupoid
\end{center}

\vspace{.1in}

\begin{center}
J.~Balog$^{(a)}$, L.~Feh\'er$^{(b)}$\protect\footnote{
Corresponding author's e-mail: lfeher@sol.cc.u-szeged.hu}
and L.~Palla$^{(c)}$ \\

\vspace{0.2in}

$^{(a)}${\em  Research Institute for Nuclear and Particle Physics,} \\
       {\em Hungarian Academy of Sciences,} \\
       {\em  H-1525 Budapest 114, P.O.B. 49, Hungary}\\

\vspace{0.2in}       

$^{(b)}${\em Institute for Theoretical Physics,
        J\'ozsef Attila University,} \\
       {\em  H-6726 Szeged, Tisza Lajos krt 84-86, Hungary }\\

\vspace{0.2in}   

$^{(c)}${\em  Institute for Theoretical Physics, 
Roland E\"otv\"os University,} \\
{\em  H-1117, Budapest, P\'azm\'any P. s\'et\'any 1 A-\'ep,  Hungary }\\

\end{center}

\vspace{.2in}

\begin{center} \bf Abstract
\end{center}

{\parindent=25pt
\narrower\smallskip\noindent
The precise relationship between the arbitrary monodromy
dependent 2-form appearing in
the chiral WZNW symplectic form and the 
\lq exchange r-matrix' that governs the corresponding  
Poisson brackets  is established.
Generalizing earlier results related to diagonal monodromy,
the exchange r-matrices are shown to satisfy a new
dynamical generalization 
of the classical modified Yang-Baxter equation,
which is found to admit 
an interpretation in terms of (new) Poisson-Lie groupoids.
Dynamical exchange r-matrices 
for which right multiplication 
yields a classical or a   
Poisson-Lie symmetry on the chiral WZNW phase space 
are presented explicitly.
}

\vspace{11 mm}
\begin{center}
keywords: WZNW model, exchange algebra, Poisson-Lie symmetry 
\end{center}
\vfill\eject

\section{Introduction}

\setcounter{equation}{0}
\renewcommand{\thefootnote}{\arabic{footnote}}
\setcounter{footnote}{0}

Since its invention fifteen years ago \cite{Witt},  
the Wess-Zumino-Novikov-Witten (WZNW)
model has been the subject of many studies, and still receives
considerable attention due  to its central role in two-dimensional
conformal field theory and to the richness of its 
structure (see e.g.~\cite{DiF}).
One of the fascinating aspects of the model is that in addition
to its built-in affine Kac-Moody symmetry
it also exhibits certain quantum group properties.
The quantum group properties were originally discovered \cite{qgroup} in
the quantized model, which raised the question
to find their classical analogues.
The intense efforts \cite{class,BDF,G,FG} at the beginning of the decade
led to the consensus that the origin of the quantum group
symmetries lies in the Poisson-Lie (P-L) -- gauge -- symmetries
of the so called chiral WZNW phase space that emerges
after splitting the left- and right-moving degrees of freedom.

However, in the definition
of the chiral WZNW symplectic structure there appears, unavoidably,
a choice to be made \cite {G}, \cite{FG}, and the 
resulting Poisson brackets (PBs)
have been analysed so far only in a small subset of the possible cases.
In this letter we wish to present new results
on the general case (for details, see \cite{LONG}).

Following \cite {G}, \cite{FG}, the chiral separation of the
space of classical solutions, $\M^{sol}$,
of the WZNW model
based on a simple Lie group $G$ with Lie algebra $\G$ can be
described as 
follows\footnote{Conventions: We have $x_L= \sigma + \tau$, 
$x_R =\sigma -\tau$,
$\pa_{C}= {\pa \over \pa x_C}$ for $C=L,R$.
For a basis $T^\alpha$ of $\G$, $I^{\alpha\beta}=\tr(T^\alpha T^\beta)$
and $[T^\alpha, T^\beta]=f^{\alpha\beta}_\gamma T^\gamma$. The dual
basis is denoted by $T_\alpha$, ${\rm Tr}(T_\alpha T^\beta
)=\delta_\alpha ^\beta $, and the elements $A\in {\cal G}$ have
components $A_\alpha ={\rm Tr}(AT_\alpha )$.
The usual summation convention is in force, indices are raised and lowered 
by $I^{\alpha\beta}$ and its inverse.
${\widetilde \G}$ contains the $2\pi$-periodic functions in 
$C^\infty({\bf R},\G)$.
}.
$\M^{sol}$  consists of the smooth $G$-valued functions
$g(\sigma, \tau)$ which are $2\pi$-periodic in the space variable $\sigma$
and satisfy the field equation 
$\pa_R (\pa_L g \cdot g^{-1})=0$.
The general solution can be written as
$g(\sigma, \tau)= g_L(x_L) g_R^{-1}(x_R)$,
where $(g_L, g_R)$ is a pair of $G$-valued, smooth,
quasiperiodic functions on the real line ${\bf R}$ 
with {\em equal monodromies}:
for $C=L,R$ one has
$g_C(x_C + 2\pi) = g_C(x_C)M$
with some $C$-independent
$M\in G$.
This means that
$\M^{sol}$ is the base of a principal $G$-bundle,
whose total space is
\be
\widehat\M= \{ (g_L, g_R) \vert g_{L,R} \in C^\infty({\bf R}, G),
\quad g_{L,R}(x + 2\pi) = g_{L,R}(x) M
\quad  M\in G\}.
\ee
The free right-action of $G$ on $\widehat \M$ is given by
$
G\ni p: (g_L, g_R) \mapsto (g_L p, g_R p)
$,
and the bundle projection,
$\vartheta: \widehat\M \rightarrow \M^{sol}$, operates as
$
\vartheta: (g_L, g_R)\mapsto g=g_L g_R^{-1}$.
Being a space of solutions, $\M^{sol}$ has a natural symplectic structure,
$\Omega_{sol}^\kappa$, whose pull-back to
$\widehat\M$ can be extended to the chirally separated space
given by $\widehat\M^{ext} = \M_L \times \M_R$ with
\be
\M_{C}= \{ g_{C} \vert g_{C} \in C^\infty({\bf R}, G),
\quad g_{C}(x + 2\pi) = g_{C}(x) M_{C}
\quad  M_{C}\in G\}.
\ee
In the extended space it is natural to require the two chiral
   factors to be completely decoupled. 
The most general symplectic structure
$\widehat \Omega_{ext}^\kappa$ on $\widehat \M^{ext}$ with this
property 
which equals $\vartheta^* (\Omega^\kappa_{sol})$ on
$\widehat \M\subset \widehat\M^{ext}$
has the form
$
\widehat \Omega_{ext}^\kappa(g_L, g_R)=
\kappa_L \Omega_{chir}^{\rho}(g_L) +\kappa_R \Omega_{chir}^{ \rho}(g_R)$.
Here  $\kappa_L=-\kappa_R = \kappa$ is a parameter 
(the affine Kac-Moody `level') and
\be
\Omega_{chir}^{\rho}(g_C) = \Omega_{chir}(g_C) + \rho(M_C)
\label{Orho}\ee
with 
\be\label{chi2form}
\Omega_{chir}(g_C)=
- {1\over 2 }\int_0^{2\pi}dx_C\,
 {\rm Tr}\left(g_C^{-1}dg_C \right) \wedge \left(g_C^{-1}dg_C \right)'
-{1\over 2} \tr \left( (g_C^{-1} dg_C)(0)\wedge dM_C\cdot M_C^{-1}\right)
\ee
and some 2-form $\rho$ depending {\em only} on the monodromy of $g_C$.
More precisely,  since in  the extended model the chiral factors
$(\M_C, \kappa_C \Omega_{chir}^{\rho})$ must be {\em symplectic}
manifolds {\em separately}, $d(\kappa_C \Omega_{chir}^{\rho})=0$,  
one has the condition \cite{G}
\be
d\rho(M_C) = {1\over 6}
\tr\left( M_C^{-1} dM_C \wedge M_C^{-1}dM_C\wedge M_C^{-1}dM_C\right),
\label{dOrho}\ee
which in turn implies that the chiral monodromy matrices $M_C$
must be restricted to a domain $\check G\subset G$
on which there exists a smooth 2-form $\rho$ satisfying (\ref{dOrho}).

One of the main points of \cite{FG} is that one can construct an 
appropriate 2-form $\rho$ out of any  (suitably normalized) antisymmetric 
solution $\hat r\in \G\wedge \G$
of the  modified classical Yang-Baxter equation (YBE)
in such a way that the corresponding PBs on $\M_C$ are encoded by the
`classical exchange algebra'
\be
\Big\{g_C(x) \stackrel{\otimes}{,} g_C(y)\Big\}
={1\over\kappa_C}\left(g_C(x)\otimes
g_C(y)\right)\left(
\hat r + {1\over 2} {\hat I} \,{\mathrm sign}\,(y-x)
\right), \quad 0< x, y<2\pi,
\label{cxchPB}\ee
where $\hat I= I^{\alpha\beta}  T_\alpha \otimes T_\beta$.
This exchange algebra admits the P-L action
$p: g_C(x) \mapsto g_C(x) p$
of the group $G=\{ p\}$
endowed with the Sklyanin bracket \cite{Skl}
\be
\kappa_C \{ p \stackrel{\otimes}{,} p\}_{\hat R} = [p\otimes p, \hat R]
\label{Sklyanin}\ee
with $\hat R=\hat r$.

By using an arbitrary $\rho$ in (\ref{Orho}), one expects to obtain  a
classical exchange algebra governed  by some monodromy dependent
`exchange r-matrix' $\hat r(M)$.
Our first  result in fact is that
we shall establish the
precise relationship between $\rho$ and $\hat r(M)$ in the general case.
We will then point out that $\hat r(M)$ and the $\hat R$-matrix
of the P-L symmetry group need not always coincide.
That is, right multiplication can be a P-L symmetry for certain
non-constant exchange r-matrices, too, which we shall give
explicitly.

To place our second  result into context, we recall
that an alternative method of the chiral separation is to
construct the general solution of the WZNW field equation out of
`Bloch-waves' that have diagonal monodromy matrix, given by
$e^\omega$ with $\omega \in \H$, where $\cal H$ is a splitting Cartan
subalgebra with basis $H^i$.
It was pointed out in \cite{BDF} that 
the corresponding monodromy dependent r-matrix,
$\hat {R}(\omega)$, then satisfies
\be
[\hat R_{12}(\omega), \hat R_{23}(\omega)] + \sum_i H_1^i  
{\pa \over \pa \omega^i}
\hat R_{23}(\omega) + \hbox{cycl.~perm.}  =-{1\over 4}\hat f,
\quad
\hat f = f_{\alpha\beta\gamma} T^\alpha\otimes T^\beta\otimes T^\beta.
\label{CDYB}\ee
This modified classical dynamical YBE appears
in other contexts \cite{Gervais,Feld,Avan}, too,
and has been much studied recently (see \cite{EV}, \cite{Liu} 
and references therein).
In \cite{EV} a geometric interpretation of (\ref{CDYB}) 
has been given in terms
of {\em dynamical P-L groupoids},
generalizing the relationship \cite{Drinfeld} between the 
(modified) classical 
YBE and P-L groups.
As our second result,
we shall show that an arbitrary exchange r-matrix of the WZNW model
is a solution of another dynamical generalization of the modified 
classical YBE, given by eq.~(\ref{GCDYB}) below, 
which also admits an interpretation in terms of 
(new) P-L groupoids.

\section{Poisson brackets on the chiral WZNW phase space}
\setcounter{equation}{0}

We here investigate the symplectic structure on the chiral 
WZNW phase space
$\M_C$ introduced above. 
The analysis is the same for both chiralities, $C=L,R$, and we 
simplify our notation by putting $\M_{chir}$ 
for $\M_C$  and $g$, $M$, $\kappa$ for $g_C$, $M_C$, $\kappa_C$, 
respectively. Thus 
  $\M_{chir}$ is parametrized by the  $G$-valued, smooth, 
quasiperiodic field $g(x)$ 
satisfying the monodromy condition 
\be 
g(x+2\pi)=g(x)M \qquad M\in G.  
\label{Moncon}\ee 
The corresponding chiral current, 
$J(x)=\kappa g^\prime(x)g^{-1}(x)\in{\cal G}$,
is a smooth, $2\pi$-periodic function of $x$. 
The domain in $\M_{chir}$ that corresponds to $M\in \check G$ is 
denoted by $\check \M_{chir}$. 
Below we show that $\kappa \Omega_{chir}^\rho$ defined by (\ref{Orho}) 
is non-degenerate if $\check G$ is appropriately chosen and describe
the main features of the PBs associated with  this symplectic form.

To use $\kappa \Omega_{chir}^\rho$ in practice we need to 
establish some notation for
tangent vectors $X[g]$ at $g\in \M_{chir}$ and vector fields $X$ over
the chiral phase space. To
this end we consider smooth curves on $\M_{chir}$ described by 
functions $\gamma(x,t)\in G$  satisfying 
 \be
\gamma(x+2\pi,t)=\gamma(x,t)M(t)\qquad
M(t)\in G;\qquad\quad \gamma(x,0)=g(x).
\label{DefMon}\ee
$X[g]$ is obtained 
as the velocity to the curve at $t=0$,
encoded by the ${\cal G}$-valued, smooth function 
\be
\xi(x):={d\over dt} g^{-1}(x)\gamma(x,t)\Big\vert_{t=0}\,.
\label{Xidef}
\ee 
The monodromy properties of $\xi (x)$ can be derived by taking 
the derivative of the first equation in (\ref{DefMon}):
$ 
\xi^\prime(x+2\pi)=M^{-1}\xi^\prime(x)M, 
$  
and this can be solved in terms of a $\G$-valued, smooth, $2\pi$-periodic 
function, $X_J\in\widetilde{\G}$, and a constant Lie algebra element,
$\xi_0$, as follows: 
\be 
\xi(x)=\xi_0+\int_0^x dy\,
g^{-1}(y)X_J(y)g(y).
\label{XipJ}
\ee
A vector field $X$ on $\M_{chir}$ is an assignment,
$g\mapsto X[g]$,
of a vector to every point $g\in \M_{chir}$.
Thus it can be specified 
by the assignments $g\mapsto \xi_0[g]\in \G$ and 
$g\mapsto X_J[g]\in\widetilde{\G}$.
Using the curve that defines $X[g]$, 
$X$ acts on a smooth function, $g\mapsto F[g]$, on $\M_{chir}$ as  
\be 
X(F)[g]={d\over d t}F[g_t]\Big\vert_{t=0}
\qquad g_t(x)=\gamma(x,t)\,.  
\ee 
Note that the evaluation functions 
$F^x[g]:=g(x)$ and ${\cal F}^x[g]:=J(x)$ are differentiable 
with respect to any 
vector field, and their derivatives are given  by 
\be
X\big(g(x)\big)= g(x)\xi(x)
\qquad\hbox{and}\qquad
X\big(J(x)\big)=\kappa X_J(x).
\label{Xsigma}\ee 
This clarifies the meaning of $X_J$ as well. 
It is also obvious from its definition that the monodromy matrix 
yields a $G$-valued differentiable function on $\M_{chir}$, 
$g\mapsto M= g^{-1}(x) g(x+2\pi )$,
whose derivative is characterized by the $\G$-valued function 
$X(M) M^{-1} =  M\xi (x +2\pi) M^{-1} -\xi (x)$.
Having defined vector fields, one can also introduce differential 
forms as usual. 
We only remark that by (\ref{Xsigma}) evaluation 1-forms like 
$dg(x)$, $dJ(x)$ or $(g^{-1}dg)^\prime(x)$ are perfectly well-defined:
e.g. $dg(x)\big( X\big)
=X(g(x))=g(x)\xi (x)$. 

The inversion of $\kappa
\Omega_{chir}^\rho$ consists of the solution of the following problem:
For a fixed (scalar) function 
$F$ on the phase space $\check \M_{chir}$, 
find the corresponding vector field, $Y^F$, satisfying 
\be 
X(F)=\kappa\Omega^{\rho}_{chir}(X,Y^F) 
\label{hamvect}\ee 
for all vector fields $X$. 
Notice that $Y^F$ does not necessarily exist for a given $F$. 
We say that $F$ is an element of the set of {\em admissible Hamiltonians},
denoted as ${\tt H}$, 
if the corresponding hamiltonian vector field, $Y^F$,  exists.
 
To compute $\kappa\Omega^{\rho}_{chir}(X,Y^F)$ we assume that 
 $X$ is parametrized by $\xi(x)$ and further by the pair 
$\big(\xi_0,X_J(x)\big)$, while   
the analogous parametrization for $Y^F$ is given  by $\eta(x)$ 
and the pair $\big(\eta_0,Y_J(x)\big)$.
Furthermore we parametrize $\rho$ now as 
\be 
\rho(M)= {1\over2}\,q^{\alpha\beta}(M){\rm Tr}
\big(T_\alpha M^{-1}dM\big)\wedge{\rm Tr}\big(T_\beta M^{-1}dM\big).
\label{qpara}\ee  
The $q^{\alpha\beta}$, $q^{\alpha\beta}=-q^{\beta\alpha}$, are smooth 
functions 
on the domain $\check G\subset G$. Studying the right hand side of
(\ref{hamvect}) one can establish \cite{LONG} the 
following three necessary and
sufficient conditions that $F$ must obey to guarantee that $Y^F$
exists:\hfill\break\noindent 
$\bullet$ There must exist 
a {\em smooth} $\G$-valued function on ${\bf R}$, 
$A^F(x)$, and a
constant Lie algebra element, $a^F$, such that for any vector field $X$ 
\be
X(F)=\kappa\int_0^{2\pi}dx \tr\Big(X_J(x)A^F(x)\Big)+ 
\kappa\tr\big(\xi_0a^F\big).
\label{Egy}\ee
This means that $F\in {\tt H}$ must have an exterior derivative 
parametrized
by the assignments $g\mapsto A^F(x)[g]$ and $g\mapsto a^F[g]$.
The restriction of $A^F(x)$ to $x\in [0,2\pi]$ and $a^F$ are
uniquely determined by (\ref{Egy}), and $A^F(x)$  is made a unique   
function on  $\bf R$ by the next requirement.  
\hfill\break\noindent $\bullet$ The expression 
\be
{1\over\kappa}\Big[A^F(x),J(x)\Big] +{A^F}^\prime(x)
\label{Ketto}
\ee
must define a smooth {\em $2\pi$-periodic} function on ${\bf R}$. 
\hfill\break\noindent $\bullet$ $A^F(x)$ and $a^F$ must be related by 
 \be
a^F=g^{-1}(0)\Big[A^F(0)-A^F(2\pi)\Big]g(0).
\label{Harom}\ee 
If these conditions are satisfied, then $Y^F$ is uniquely determined,
and is in fact given by  
\be
g^{-1}(x) Y^F(g(x))= g^{-1}(x)A^F(x)g(x)-{1\over2}a^F+r(M)\big(a^F\big)\,,
\label{etasol}
\ee 
where 
\be 
r(M)\big(a^F\big)= T_\alpha r^{\alpha\beta}(M) a^F_\beta 
\label{RM}\ee 
and the matrix $r^{\alpha\beta}(M)$ is defined as the solution 
of an interesting linear equation. This equation can be best described
in terms of the linear operators $q_\pm (M)$, 
$r_\pm (M)$ $\in {\mathrm End}(\G)$
which are associated with the matrices 
\be
r_\pm^{\alpha\beta}(M)=r^{\alpha\beta}(M) \pm {1\over 2} I^{\alpha\beta}
\quad\hbox{and}\quad
q_\pm^{\alpha\beta}(M)=q^{\alpha\beta}(M) \pm {1\over 2} I^{\alpha\beta} 
\label{rqpm}\ee 
by the rule  
\be\label{rule}
Q(A)= T_\alpha Q^{\alpha\beta} A_\beta,
\qquad
\qquad Q^{\alpha\beta}=r_\pm^{\alpha\beta}(M),\
q_\pm^{\alpha\beta}(M).
\ee
In fact \cite{LONG},  using these operators the equation determining
$r^{\alpha\beta}(M)$ obtains the form:
\be
q_+(M)\circ r_-(M)= q_-(M)\circ \Ad (M^{-1}) \circ r_+(M).
\label{requfact}\ee
Notice that at the identity $e\in G$ the unique solution is 
$r^{\alpha\beta}(e)=q^{\alpha\beta}(e)$. 
It follows by continuity  that there is a neighbourhood of $e\in G$ 
 on which a unique solution exists and is a smooth function of $M$.
   Moreover, it is easy to see that the solution $r(M)$ 
is antisymmetric,  
$r^{\alpha\beta}=-r^{\beta\alpha}$.
By choosing $\check G\subset G$ appropriately, hence we obtain 
by (\ref{requfact})   
a {\em one-to-one} correspondence between the smooth 2-form 
$\rho$ on $\check G$
and the smooth map $\check G\ni M\mapsto r(M)\in {\mathrm End}(\G)$.

Incidentally,  it follows from  (\ref{etasol}) that  
\be
Y^F (J(x))= \Big[A^F(x),J(x)\Big]+\kappa {A^F}^\prime(x),
\label{YJ}\ee   
which explains why (\ref{Ketto}) must define an element of 
$\widetilde{\G}$.

Having obtained the conditions of existence and the general
expression (\ref{etasol}) for hamiltonian vector fields, one can 
analyse which
functions belong to ${\tt H}$. 
First note that  the evaluation
functions ${\cal F}^x[g]$ and $F^x[g]$  fail to satisfy the first
condition, thus they are not in ${\tt H}$. However, their smeared out
versions
 \be\label{smeardef} 
\F_\mu:=\int_0^{2\pi}dx{\tr}\Big(\mu(x)J(x)\Big),\qquad 
F_\phi[g]:=\int_0^{2\pi}dx {\tr}\Big(\phi(x)g^\Lambda(x)\Big), 
\ee
(where in defining $F_\phi$ we use a representation $\Lambda:
G\rightarrow GL(V)$ of $G$ with $g^\Lambda =\Lambda (g)$ and 
  a smooth test function $\phi: \R \rightarrow {\mathrm End}(V)$) can be
 shown to belong to ${\tt H}$, if 
$\mu(x)$ is a $\G$-valued, smooth, $2\pi$-periodic test function, and
$\phi $ satisfies $\phi^{(k)}(0)=\phi^{(k)}(2\pi)=0$ for every 
integer $k\geq 0$. 
The corresponding hamiltonian vector fields obtained from (\ref{etasol}) 
satisfy 
\be
Y^{\F_\mu}\big(g(x)\big)=\mu(x)g(x),
\quad
Y^{\F_\mu}(J(x)) = [\mu(x), J(x)] + \kappa \mu'(x),
\qquad
Y^{\F_\mu}(M)=0, 
\label{YofJ}\ee
and, for $x\in [0,2\pi]$,  
\be
g^{-1}(x) Y^{F_\phi}(g(x))= 
{1\over\kappa} T^\alpha  
\int_x^{2\pi}d y \tr\left( T^\Lambda_\alpha \phi(y) g^\Lambda(y)\right)
 -{1\over 2} a^{F_\phi} + r(M)(a^{F_\phi}).
\label{Yg}\ee
Eq.~({\ref{YofJ}) shows that the $\F_\mu$ generate an infinitesimal action 
of the loop group on the phase space with respect to which $g(x)$ is an 
affine Kac-Moody primary field, and the current $J(x)$ 
transforms according to the co-adjoint action of the  
centrally extended loop group.
The matrix elements $M^\Lambda_{kl}$ of 
the monodromy matrix in representation
$\Lambda $ also belong to ${\tt H}$.
The action of $Y^{M^\Lambda_{kl}}$ on
$g_{ij}^\Lambda (x)$ and on $M_{ij}^\Lambda $ can be written in tensorial
form as
 \be
Y^{M^\Lambda_{kl}}\bigl( g^\Lambda_{ij}(x)\bigr)
=
{1\over \kappa} \bigl( g(x)\otimes M \, 
\hat \Theta(M)\bigr)^\Lambda_{ik, jl}\,,\label{gMPB}\ee
\be
 Y^{M^\Lambda_{kl}}(M^\Lambda_{ij}) = {1\over \kappa} \Big( (M\otimes M) 
\hat \Delta(M)  \Big)^\Lambda_{ik,jl}\,,
\label{MMPB}\ee
where 
our tensor product notation is $(K\otimes L)_{ik,jl}= K_{ij}
L_{kl}$, and 
\be
\hat \Theta(M) = \hat r_+(M)  -M_2^{-1} \hat r_-(M) M_2\,,\qquad
\hat \Delta(M) = \hat \Theta(M) - M_1^{-1} \hat \Theta(M) M_1 
\label{Delta}\ee 
with $\hat r_\pm (M)=r_{\pm }^{\alpha\beta}(M)T_\alpha\otimes T_\beta$.
The matrix $r^{\alpha\beta}(M)$ appearing here is the solution of
(\ref{requfact}), and $M_1= M\otimes 1$, $M_2= 1\otimes M$. 

We now wish to interpret the above hamiltonian vector fields 
in terms of PBs. 
Recall that the PB of two smooth functions $F_1$ and $F_2$ 
on a {\em finite dimensional} smooth symplectic manifold is defined by 
\be
\{ F_1, F_2\} = Y^{F_2}(F_1)=-Y^{F_1}(F_2)=\Omega(Y^{F_2},Y^{F_1}),
\ee
where $Y^{F_i}$ is the hamiltonian vector field associated with $F_i$
by the symplectic form $\Omega$.
One may formally apply the same formula in the infinite dimensional
case to the admissible functions that possess a hamiltonian vector field.
However, it is then a non-trivial problem to fully specify the
set of functions that form a closed Poisson algebra.
Setting this question aside, 
it is clear from (\ref{YofJ}) and (\ref{MMPB}) 
that the admissible functions of $J$ and those of $M$ 
will form two closed Poisson subalgebras that centralize each other.  
Furthermore, we may  
use the perfectly well-defined expression 
\be
\{ F_\chi, F_\phi\}:= Y^{F_\phi}( F_\chi)
\ee
for the PB of two admissible Hamiltonians  
of  type $F$ in eq.~(\ref{smeardef}) to define the 
(`distribution valued') PB of
the evaluation functions $g(x)$ by the equality:
\be
 \{ F_\chi, F_\phi\}:=
\int_0^{2\pi} \int_0^{2\pi} dx dy  {\mathrm Tr}_{12}
\left( \chi(x)\otimes \phi(y) 
\{ g^\Lambda(x) \stackrel{\otimes}{,} g^\Lambda(y)\}\right),
\label{localPB}\ee
where ${\mathrm Tr}_{12}$ is the (normalized) trace over $V\otimes V$ and 
$
\{ g^\Lambda(x) \stackrel{\otimes}{,} g^\Lambda(y)\}_{ik,jl}= 
\{ g^\Lambda_{ij}(x), g^\Lambda_{kl}(y)\}$.   
With these definitions, our explicit formula of the hamiltonian vector
field $Y^{F_\phi}$ in (\ref{Yg})
is equivalent to the following quadratic exchange algebra type
PB for the chiral field $g(x)$:
\be 
\Big\{g^\Lambda(x)\stackrel{\otimes}{,} g^\Lambda(y)\Big\}
={1\over\kappa}\Big(g^\Lambda(x)\otimes
g^\Lambda(y)\Big)\Big(
\hat r(M) + {1\over 2} {\hat I} \,{\mathrm sign}\,(y-x)
\Big)^\Lambda, \quad 0< x, y<2\pi,
\label{xchPB}\ee  
where 
$\hat r(M)$  denotes (the appropriate representation of)
the element in ${\cal G}\otimes{\cal G}$ corresponding to the solution
of eq.~(\ref{requfact}): $\hat r(M)=r^{\alpha\beta}(M)T_\alpha\otimes
T_\beta$. 
Thus eq.~(\ref{requfact}) gives indeed the precise relation between 
the 2-form $\rho$ in (\ref{Orho}), (\ref{qpara}) 
and the -- in general monodromy dependent --
`exchange r-matrix', $\hat r(M)$. 
 Proceeding in the same way with
the $\{ F_\phi , M_{kl}^\Lambda\}$ PB
 as we did with the $\{ F_\chi,
F_\phi\}$ one, we conclude that the right hand side of 
(\ref{gMPB}) should be interpreted as the expression of
the $\{ g_{ij}^\Lambda (x) ,M_{kl}^\Lambda\}$ PB,
and similarly for (\ref{MMPB}).

It is clear from the foregoing discussion that 
the admissible Hamiltonians of type ${\cal F}_\mu$, $F_\phi$ and 
$M^\Lambda_{kl}$ should together generate a 
closed Poisson algebra. 
Although at present we cannot 
fully characterize the set of elements that belong to this algebra,
we wish to point out that the Jacobi 
identity for three functions of type $F_\phi$,
in any Poisson algebra that contains them, 
is equivalent  to the following equation for $\hat r(M)$:
\be
\big[\hat r_{12}(M),\hat r_{23}(M)\big]
+\Theta_{\alpha\beta}(M)T^\alpha_1{\cal R}^\beta \hat r_{23}(M)
+ \hbox{cycl. perm.}=
-{1\over 4} \hat f.
\label{GCDYB}
\ee
Here $\hat f$ is the same as in (\ref{CDYB}) 
and the cyclic permutation is over the three tensorial factors
with  
$\hat r_{23} = r^{\alpha\beta}  (1\otimes T_\alpha\otimes T_\beta)$,
$T^\alpha_1= T^\alpha \otimes 1 \otimes 1$ and so on.
Furthermore, we use the components of  
$\hat \Theta=\Theta_{\alpha\beta}T^\alpha\otimes T^\beta$
given by (\ref{Delta}), and the left-invariant differential
operators ${\cal R}^\beta$ that act on a function $\psi$ of $M$ by 
\be
({\cal R}_\alpha \psi)(M):= {d\over d t}
 \psi(Me^{t T_\alpha} )\Big\vert_{t=0},
\qquad 
{\cal R}^\beta = I^{\beta \alpha} {\cal R}_\alpha.
\label{derLR}\ee
Eq.~(\ref{GCDYB}) can be viewed as a dynamical generalization of the 
classical modified YBE, to which it reduces 
if the r-matrix is a monodromy independent constant.
Of course, (\ref{GCDYB}) is satisfied for any $\hat r(M)$ that arises
as a solution of (\ref{requfact}) since the Jacobi identity is guaranteed
by $d\Omega^{\rho}_{chir}=0$.

Next we describe an interesting solution of (\ref{GCDYB}) obtained 
by inverting
(\ref{requfact}) for the r-matrix using a particular 2-form 
$\rho$ (\ref{qpara}) as input.
For this, we now note that  
if the monodromy matrix $M$ is near to $e\in G$, 
then the chiral WZNW field can be uniquely parametrized as
\be
g(x) = h(x) e^{x\Gamma},
\label{expara}\ee
where $h(x)$ is a $G$-valued, smooth, $2\pi$-periodic function
and $\Gamma$ varies in a neighbourhood of zero in $\G$,
$\check \G\subset \G$, for which the map 
$\check \G \ni\Gamma \mapsto M=e^{2\pi \Gamma} \in \check G$ 
is a diffeomorhism. 
An easy computation gives the following formula for 
$\Omega_{chir}$ (\ref{chi2form}) in this parametrization of 
$\check \M_{chir}$: 
\be
\Omega_{chir} (h,\Gamma)= 
\Omega^0_{chir} (h, \Gamma) -\rho_0(\Gamma), 
\label{Omegapar}\ee
where
 \be
\Omega^0_{chir} (h,\Gamma )=
-{1\over 2}\int_0^{2\pi}dx\, \tr\left(h^{-1} dh 
\wedge (h^{-1} dh )'\right)
+d \int_0^{2\pi}dx \tr\left(\Gamma h^{-1} dh\right) ,
\label{Omega0}\ee
\be
\rho_0(\Gamma) = -{1\over 2}\int_0^{2\pi}dx\, 
\tr\left(d \Gamma \wedge de^{x \Gamma} e^{-x \Gamma}\right).
\label{rho0}\ee
Taking into account that $M=e^{2\pi \Gamma}$,
it is not difficult to verify that
\be
d\Omega^0_{chir}=0,\qquad 
 d\rho_0(\Gamma) = {1\over 6} \tr
\left( M^{-1}dM\wedge M^{-1}dM\wedge M^{-1}dM\right).
\label{drho0}\ee
Recalling eq.~(\ref{dOrho}), we see that the 2-form $\rho $ in (\ref{Orho})  
in this case can be 
parametrized by an  arbitrary closed 2-form
$\beta$ on $\check \G$ as 
\be
\rho(\Gamma)=\rho_0(\Gamma) + \beta(\Gamma),\qquad   
d\beta(\Gamma)=0. \quad 
\label{beta}\ee
By (\ref{Orho}) we thus have 
$\Omega^\rho_{chir}= \Omega_{chir}^0 + \beta$, 
in particular  $\Omega^{\rho_0}_{chir}=\Omega^0_{chir}$. 
In order to find the exchange r-matrix, $\hat r_0$,  
corresponding to $\rho_0$,
we notice that the integral defining $\rho_0$ can be computed in
closed form and the linear operator, $q_0$,  corresponding 
by (\ref{rule}) to its
matrix is given by
\be
q_0=\frac{2{\cal Y}+e^{-\cal Y}-e^{\cal Y}}
{2(e^{\cal Y}-1)(1-e^{-\cal Y})}
\qquad \hbox{with}\qquad
{\cal Y}:=2\pi ({\rm ad}\,\Gamma ).
\label{qexpl}\ee
Then from eq.~(\ref{requfact}) we find the linear 
operator\footnote{The expressions in eqs.~(\ref{qexpl}), 
(\ref{ralt}), (\ref{rveg})
are defined by the power series expansions of the corresponding complex
analytic functions around zero. For instance \cite{GR},
$2r_0= \sum_{k=1}^\infty { 2^{2k}  B_{2k} \over (2k)!} 
({1\over 2} {\cal Y})^{2k-1}$.} 
version, $r_0$,  of the
exchange r-matrix as
\be
r_0=\frac{1}{2}\coth\frac{ {\cal Y}}{2}-\frac{1}{{\cal Y}}\,.
\label{ralt}\ee
By means of  (\ref{xchPB}) this r-matrix 
defines one of the possible monodromy
dependent exchange algebras for the chiral WZWN field,
and it also represents a non-trivial special solution of (\ref{GCDYB}). 
To uncover a remarkable property of this solution,
we note that the function $F(h,\Gamma)=\Gamma_\alpha=\tr(T_\alpha
\Gamma)$ belongs to $\tt H$, and its hamiltonian vector field
in the $\rho=\rho_0$ case gives rise to the PBs: 
 \be
\{ g(x), \bar \Gamma_\alpha\} = g(x) T_\alpha
\quad\hbox{and}\quad 
\{ \bar \Gamma_\alpha, \bar \Gamma_\beta\} = -
f_{\alpha\beta}^{\gamma} \bar \Gamma_\gamma
\quad
\hbox{for}\quad 
\bar\Gamma_\alpha:= 2\pi \kappa \Gamma_\alpha\,.  
\ee 
This means that  in the case of the symplectic form 
$\kappa\Omega_{chir}^0$ the $\bar \Gamma_\alpha$ (essentially the
logarithm of the monodromy matrix) generate a  {\em classical}
${\cal G}$-symmetry  on $\check {\cal M}_{chir}$. 
We remark in passing that a classical $\G$-symmetry 
is sometimes called
`Abelian' to contrast it with a proper (`non-Abelian') P-L symmetry,
for which the symmetry group itself is endowed with a non-zero PB
(for P-L symmetry, see e.g.~\cite{FG}, \cite{BB} and references therein). 

Motivated by the somewhat surprising result obtained above, 
we now investigate what
conditions the monodromy dependent exchange r-matrix should satisfy 
in general to
guarantee that the standard (rigid) right action of $G$ on 
$\check {\cal M}_{chir}$ is a
proper P-L symmetry.  
For this purpose we endow the group with the Sklyanin bracket      
(\ref{Sklyanin}) (replacing $\kappa_C$ now by $\kappa$), 
 where $\hat R=R^{\alpha\beta}T_\alpha\otimes T_\beta\in \G\wedge \G$
 is a 
{\em constant} r-matrix subject to the requirement 
\be
[\hat R_{12}, \hat R_{23}] +\hbox{cycl. perm.} = -\nu^2 \hat f
\qquad \nu^2 : \hbox{some real constant}.
\label{nnu}\ee
It is easy to check that  the rigid right 
action\footnote{Since $M\mapsto p^{-1}Mp$,  strictly speaking  
we here have to assume that $\check G\subset G$ is invariant under 
the adjoint action of $G$, or else  
 the statements should be reformulated in terms of the corresponding 
$\G$-action.}  
of $G$ on $\check \M_{chir}$,
$
p: g(x) \mapsto g(x) p
$ $\forall p\in G$,
is a P-L action if and only if
\be
\hat r(p^{-1}Mp)-\hat R = (p\otimes p)^{-1} 
\bigl( \hat r(M) -\hat R\bigr) (p\otimes p).
\label{PLcondition}\ee
This simply means that the $\G\wedge \G$-valued function
 $(\hat r(M)- \hat R)$ on $\check G$ must be {\em equivariant}
with respect to the natural (infinitesimal)  
actions of $G$ on $\check G$ and on $\G \wedge \G$.

Therefore the right multiplication is a P-L symmetry  iff 
the exchange r-matrix $\hat r(M)$ is such a solution of
(\ref{GCDYB}) that the corresponding difference $(\hat r(M)- \hat R)$ is
equivariant. Insisting on the parametrization $M=e^{2\pi \Gamma}$, 
the search for these solutions is made feasible by the observation that
any analytic function of ${\cal Y}=2\pi (\ad{\Gamma})$ 
is equivariant.
In fact, 
one of our main results, proved in \cite{LONG}, 
is that  the r-matrix corresponding to the linear operator
\be
r=\frac{1}{2}\coth\frac{ {\cal Y}}{2}-\nu\coth(\nu{\cal Y})+R 
\label{rveg}\ee
solves (\ref{GCDYB}).
Some remarks on this formula are in order.
First, note that in the $\nu=0$ case we mean the limit of the 
corresponding
complex analytic function, whereby we  
recover $r_0$ in (\ref{ralt}) if in addition we use $R=0$ 
(see also footnote 2).
Second, notice that if  $\nu =\frac{1}{2}$, then $r=R$,
which is the case of the constant exchange r-matrices \cite{FG}.
Third,  it is worth stressing that for a compact Lie algebra $\G$
constant exchange r-matrices do not exist,  
because of the negative sign on the right hand side of (\ref{GCDYB}),
but our formula (\ref{rveg}) gives explicit solutions also in this case
using a purely imaginary $\nu$ in (\ref{nnu}).
Finally, we remark  that in the $\nu={1\over 2}$ case 
the construction of the 2-form $\rho$ that corresponds 
to the r-matrix in (\ref{rveg}) is presented in \cite{FG}, while 
in general the existence (and the uniqueness) of a suitable 
local 2-form is guaranteed by the
solvability of (\ref{requfact}).
Further comments and explicit results are contained in \cite{LONG}.

\section{Encoding chiral WZNW by Poisson-Lie groupoids}
\setcounter{equation}{0}

The classical dynamical YBE (\ref{CDYB}) can be 
regarded as the guarantee of the Jacobi identity in
a P-L groupoid \cite{EV}. 
Below we show that eq.~(\ref{GCDYB}) 
admits  an analogous interpretation.
For this, we introduce a family of 
P-L groupoids in such a way that a subfamily 
of these is naturally associated with the possible 
PBs on the chiral WZNW phase space.
Remarkably, these groupoids are finite dimensional Poisson manifolds  
that encode practically all information about the
infinite dimensional chiral WZNW PBs.

Roughly speaking, a groupoid is a set, say $P$,  
endowed with a `partial multiplication'
that behaves similarly to a group multiplication in 
the cases when it can be performed.
To understand the following construction one does not 
need to know details of the notion of a groupoid
(see e.g.~\cite{Gpoid}), since we shall only use the most 
trivial example of such a structure, for which 
\be
P= S \times G \times S = \{ (M^F, g, M^I) \},
\label{Pset}\ee 
where $G$ is a group and $S$ is some set. 
The partial multiplication is defined for those triples 
$(M^F, g, M^I)$ and $(\bar M^F, \bar g, \bar M^I)$ for which 
$M^I=\bar M^F$,
and the product is 
\be
(M^F, g, M^I) (\bar M^F, \bar g, \bar M^I):= (M^F, g\bar g, \bar M^I)
\quad\hbox{for}\quad M^I=\bar M^F.
\label{Pmult}\ee 
In other words, the graph of the partial multiplication
is the subset of 
\be
P\times P\times P =
\{ (M^F, g, M^I)\} \times \{ (\bar M^F, \bar g, \bar M^I)\}
\times \{ (\hat M^F, \hat g, \hat M^I)\} 
\ee
defined by the constraints
\be
M^I=\bar M^F,
\quad
\hat M^F= M^F,
\quad
\hat M^I=\bar M^I,
\quad
\hat g= g\bar g,
\label{graph}\ee
where the hatted triple encodes the components of the product. 
A P-L groupoid \cite{Wei} $P$ is a groupoid and a Poisson manifold 
in such a way that the graph of the partial
multiplication is a {\em coisotropic} submanifold of  
$P\times P\times P^-$, where $P^-$ denotes the  manifold
$P$ endowed with the opposite of the PB on  $P$.
In other words, this means that {\em the constraints that define 
the graph are first class}. 
This definition 
reduces to that of a P-L group in the particular 
case for which the set $S$ in (\ref{Pset}) consists of a single point.

In the interpretation of (\ref{CDYB}) given in \cite{EV} the groupoid 
$P$ is of the form above with $S$ taken to be a domain in 
the dual of a Cartan subalgebra of a simple Lie group $G$.
By thinking about a generic monodromy matrix, 
we  now take $P$ to be
\be
P= \check G \times G \times \check G,
\ee
where $\check G$ is some open domain in $G$.
On this $P$, we postulate a PB  $\{\ ,\ \}_P$ defined, 
by using the usual tensorial notation, as follows: 
\bea
&&
\kappa \{ g_1, g_2\}_P = g_1 g_2 \hat r(M^I) - 
\hat r(M^F) g_1 g_2 
\nonumber\\
&& \kappa \{ g_1, M^I_2\}_P = g_1 M_2^I \hat \Theta(M^I)
\nonumber\\
&& \kappa \{ g_1, M_2^F\}_P = M_2^F \hat\Theta(M^F) g_1
\nonumber\\
&&\kappa \{ M^I_1, M^I_2\}_P = M^I_1 M^I_2 \hat\Delta(M^I)
\nonumber\\
&&\kappa \{ M^F_1, M^F_2\}_P = - M^F_1 M^F_2 \hat\Delta(M^F)
\nonumber\\
&&\kappa \{ M^I_1, M^F_2\}_P =0.
\label{groupoidPB}
\eea
Here $\kappa$ is an arbitrary constant included for comparison purposes.
The `structure functions' $\hat r$, $\hat \Theta$, $\hat \Delta$ 
are $\G\otimes \G$ valued functions 
on $\check G$; in components 
\be
\hat r(M)= r^{\alpha\beta}(M) T_\alpha\otimes T_\beta,
\quad
\hat \Theta(M)= \Theta^{\alpha\beta}(M) T_\alpha\otimes T_\beta,
\quad
\hat \Delta(M)= \Delta^{\alpha\beta}(M) T_\alpha\otimes T_\beta.
\ee 
It is quite easy to verify that a PB given by
the ansatz (\ref{groupoidPB}) 
always yields a P-L groupoid, since the constraints
in (\ref{graph}) will be first class for any choice
of the structure functions.
Of course, the structure functions must satisfy a system
of equations in order for the above ansatz to define a PB.
The antisymmetry of the PB is ensured by
\be 
\hat r=-\hat r_{21}
\quad (\hat r_{21}:= r^{\alpha\beta} T_\beta \otimes T_\alpha)
\qquad\hbox{and}\qquad \hat 
\Delta=-\hat \Delta_{21},
\ee
while  the Jacobi identity is, in fact, equivalent to the following
system of equations:
\bea
&& 
[ \hat r_{12}, \hat r_{13}] + \Theta_{\alpha\beta}
T^\alpha_1 {\cal R}^\beta \hat r_{23} + \hbox{cycl. perm.}= \mu \hat f ,
\quad \mu=\hbox{constant},
\label{J1}\\
&& 
[ \hat \Delta_{12}, \hat \Delta_{13}] + 
\Delta_{\alpha\beta} T^\alpha_1 {\cal R}^\beta \hat \Delta_{23}
+ \hbox{cycl. perm.}=0, 
\label{J4}\\
&& [\hat r_{12}, \hat \Theta_{13}+\hat \Theta_{23}] + 
[\hat \Theta_{13}, \hat \Theta_{23}] 
+\Delta_{\alpha\beta} T^\alpha_3 {\cal R}^\beta \hat r_{12} 
+\Theta_{\alpha\beta}( T^\alpha_1 {\cal R}^\beta \hat \Theta_{23}
- T^\alpha_2 {\cal R}^\beta \hat \Theta_{13})=0,\qquad\qquad
\label{J2}\\
&& [\hat\Theta_{12}+\hat\Theta_{13}, \hat\Delta_{23}] +
[\hat\Theta_{12},\hat \Theta_{13}]
+\Theta_{\alpha\beta} T^\alpha_1 {\cal R}^\beta \hat\Delta_{23}
+\Delta_{\alpha\beta} (T^\alpha_3 {\cal R}^\beta \hat \Theta_{12} 
-T^\alpha_2 {\cal R}^\beta \hat\Theta_{13})=0.\qquad
\label{J3}\eea
Observe that the left hand side of (\ref{J1}) 
is of the same form as that of (\ref{GCDYB}), but in the groupoid context
on the right hand side we have an {\em arbitrary} constant $\mu$.
The derivation of the above equations from the various instances of 
the Jacobi identity is not difficult. What is somewhat miraculous 
is that one does not obtain more equations than these.
This is actually ensured by our choice of the relationship between
the PBs that involve $M^I$ and those that involve $M^F$.
As an illustration, let us explain how 
(\ref{J1}) is derived.
By evaluating 
\be
\{ \{g_1, g_2\}_P, g_3\}_P + \hbox{cycl. perm.}=0,
\ee
one obtains that this is equivalent to 
\bea
&&g_1 g_2 g_3 
\left([ \hat r_{12}, \hat r_{13}] + \Theta_{\alpha\beta}
T^\alpha_1 {\cal R}^\beta \hat r_{23} + \hbox{cycl. perm.}\right)(M^I)=
\nonumber \\
&&\qquad =
\left([ \hat r_{12}, \hat r_{13}] + \Theta_{\alpha\beta}
T^\alpha_1 {\cal R}^\beta \hat r_{23} + 
\hbox{cycl. perm.}\right)(M^F) g_1 g_2 g_3.
\eea
This holds if and only if the expression in the parenthesis
is a constant, ${\mathrm Ad}$-invariant element of $\wedge^3(\G)$,
and $\mu \hat f$ is the only such element for a simple Lie algebra $\G$.

We have seen that the chiral WZNW  
PBs are encoded by equations (\ref{xchPB}), (\ref{gMPB}) and
(\ref{MMPB}), where $\hat \Theta$ and $\hat \Delta$ are defined 
by (\ref{Delta}) 
in terms of a solution $\hat r$ of (\ref{GCDYB}).
Now our point is the following: {\em A P-L groupoid can be 
naturally associated 
with any Poisson structure on the chiral WZNW phase space 
by taking the triple $\hat r$, $\hat \Theta$, $\hat \Delta$
that arises in the WZNW model to be the structure functions of a 
P-L groupoid 
according to (\ref{groupoidPB}).}

It can be checked that the Jacobi identities of the P-L groupoid
(\ref{J1})--(\ref{J3}) are satisfied for any triple 
$\hat r$, $\hat \Theta$, $\hat \Delta$ that arises in the WZNW model. 
This actually follows without any computation since, indeed, 
the Jacobi identities of the chiral WZNW PBs in (\ref{xchPB}), 
(\ref{gMPB}),
(\ref{MMPB}) lead to the same equations, with $\mu=-{1\over 4}$, 
and they are satisfied since they follow from the 
symplectic form $\kappa \Omega_{chir}^\rho$.   

Among the `chiral WZNW P-L groupoids' described above there are 
those special cases for which $\hat r$ satisfies 
(\ref{PLcondition}) with some constant r-matrix $\hat R$   
in correspondence with a right P-L action 
of $G$  on  the chiral WZNW phase space.
Under this circumstance one can verify that the two maps defined by
\be
P \times G \ni \bigl( (M^F, g, M^I), p \bigr) \mapsto (M^F, gp, 
p^{-1} M^I p)\in P
\label{rightaction}\ee
and respectively by
\be
G \times P\ni \bigl( p, (M^F, g, M^I)\bigr) \mapsto (pM^F p^{-1}, 
pg, M^I)\in P
\label{leftaction}\ee
are {\em both}  Poisson maps.
Thus they define (see also footnote 3) 
respectively {\em a right and a left} P-L action 
of the P-L group $G$, endowed with the PB (\ref{Sklyanin}) 
for $\kappa_C=\kappa$, 
on the P-L groupoid $P$.

In \cite{EV} P-L groupoids are associated with arbitrary subalgebras 
${\cal K}\subset \G$, although the corresponding dynamical r-matrices 
are described only if ${\cal K}$ is a Cartan subalgebra.
The ${\cal K}=\G$ special case of their 
groupoids  is in fact equivalent to our P-L groupoid whose structure 
function is the r-matrix in (\ref{ralt}).
Their P-L groupoids are different from ours in general.

\medskip

The reader may find a detailed exposition of the 
subject of this letter  and  several related issues in \cite{LONG}. 
Among the questions for future study,
it would be interesting to investigate the quantization  
of the above introduced P-L groupoids and to find 
other applications for them in the field of integrable systems.
As for the second question,  recall that
the classical dynamical YBE  appears
not only in the classical chiral WZNW model, but also 
in the theory of the Knizhnik-Zamolodchikov-Bernard 
equation \cite{Feld}, the Calogero-Moser 
systems \cite{Avan} etc, and thus perhaps it might be natural to
ask if (\ref{GCDYB}) and the associated P-L groupoids 
can have other interesting  applications.

\bigskip
\noindent
{\bf Acknowledgements.}
This investigation was supported in part by the Hungarian National
Science Fund (OTKA) under T019917, T030099, T025120 
and by the Ministry
of Education under FKFP 0178/1999 and FKFP 0596/1999.

\end{document}